\newcommand{\mper}{\mbox{\ \ .} }
\newcommand{\mcomma}{\mbox{\ \ ,} }
\documentstyle[11pt,psfig]{article}
\title{
On the equality of Hausdorff and box counting dimensions
}
\date{September 1990 }
\author{
		Ronnie Mainieri
		\\
		{\em Center for Nonlinear Studies, MS B258} \\
		{\em Los Alamos National Laboratory, } \\
		{\em Los Alamos, NM 87545}
}

\begin{document}

\maketitle

\begin{abstract}
%%BeginAbstract
By viewing the covers of a fractal as a statistical mechanical
system, the exact capacity of a multifractal is computed. The
procedure can be extended to any multifractal described by a
scaling function to show why the capacity and Hausdorff
dimension are expected to be equal. 
%%EndAbstract
\end{abstract}

\section{Introduction}

The strange sets that arise in the study of dynamical systems
are difficult to characterize and to compare with theoretical
predictions.  A simple test to check if two strange sets are not
smooth distortions of each other is to compare their fractal
dimensions: if the fractal dimensions differ, then the sets are
not related. There are two widely used definitions of fractal
dimensions: the Hausdorff dimension and the capacity (or box
counting) dimension.  They have long been conjectured 
\cite{FarmerDimension,TakensReconstruction,ChicagoFofa}
to be numerically equal for fractals generated by smooth
dynamical systems, even though they are not equal in general
\cite{FarmerDimension}.

The Hausdorff dimension and the capacity dimension have similar
definitions.  In both cases we measure how the sum of the boxes in
a cover of the fractal scales as the largest box is taken to
zero, but in the Hausdorff case the sum is minimized by allowing
different box sizes.  It is the minimization that gives the
Hausdorff dimension its theoretical advantage, as it excludes
pathologies that may arise in the limit of smaller boxes and
countably many isolated points. Due to the minimization procedure
in its definition, the Hausdorff dimension is invariant under
diffeomorphisms, while the capacity is invariant under a more
restrictive set of transformations:  those that transform the
metric into an equivalent metric \cite{YorkeDimensionInvariance}.

The Hausdorff dimension has a larger theoretical interest, but it
is capacity like dimensions that are usually determined
experimentally, as current algorithms make no attempt to optimize
the coverings. The now standard procedure of Packard {\em et al{.}\/}
\cite{FarmerReconstruction} and Takens \cite{TakensReconstruction}
reconstructs the attractor up to arbitrary coordinate
transformations under which the capacity may not be invariant
(see examples in \cite{YorkeDimensionInvariance}).

Here we show that the two dimensions agree for fractals that are
described by rapidly convergent scaling functions.  The scaling
function relates successive refinements of covers of the fractal
which can be organized in a tree, the branches indicating how a
subcover was refined.  The different branches of the tree are then
considered as different configurations of a canonical ensemble
where the energy of the configuration is proportional to the size
of the subcover represented by the branch.  Scaling functions and
related concepts have received little discussion in the physics
literature, so I will present the basic ideas through a simple
example.  In section 2 we use the scaling function to describe a
four scale Cantor set and show how this description is related to
a presentation function, or its inverse, a set of contractions that
generate the same set.  Different ways of organizing the tree
correspond to different ways of describing the same set. The capacity
is related to the microcanonical ensemble and the corresponding
tree is denoted a capacity tree.  This notion turns out to be useful
in analyzing box counting concepts such as lacunarity, local scaling,
and convergence rates, not all explored here.  In section 2 we also
exactly compute the capacity of a multifractal with equiprobable
boxes.  The methods of section 2 will be generalized to arbitrary
fractals described by scaling functions in section 3 where the
equality of the two dimensions is established.

\section{Capacity of a multifractal}

Most of the fractals that arise in the study of physical systems cannot
be described by a simple number, such as the fractal dimension.  In a
simple deterministic fractal any of its part is an exact rescaling of
the whole, but this is not always the case.  As observed by Frisch and
Parisi \cite{ParisiMultifractal}, different parts of the fractal have
to be rescaled by different amounts to resemble the whole, and they are
better described by a multitude of scales --- they are multifractals.
One way of describing a multifractal is in analogy with the
middle-third Cantor set: by a sequence of covers that converge to the
points of the set.  The covers can be described by a subtraction
processes, where the middle third of every cover is removed to generate
a new set of covers.  They can also be described by a contraction
processes, where the covers at one level get contracted and translated
to generate the new set of covers.  The contraction processes is
not as well known as the subtraction processes, but it is closer
in spirit to the way fractals are generated in dynamical
systems.

\begin{figure}
  \centerline{\psfig{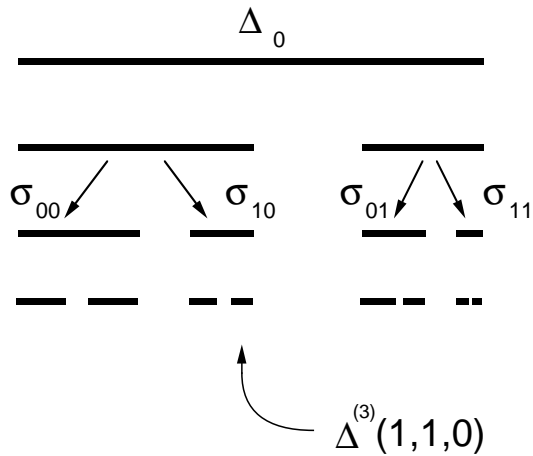}}
  \caption{\protect\footnotesize Covers of the four scale Cantor set}
  \label{fig1}
\end{figure}

A simple example of a multiscale fractal on a binary tree is the
four-scale Cantor set. We start with two initial segments, the initial
conditions $\Delta^{(1)}(0)$ and $\Delta^{(1)}(1)$. The segment
$\Delta^{(1)}(0)$ is contracted by $\sigma_{00}$ and $\sigma_{10}$
producing two new segments $\Delta^{(2)}(0,0)$ and $\Delta^{(2)}(1,0)$;
and the segment $\Delta^{(1)}(1)$ is contracted by $\sigma_{01}$ and
$\sigma_{11}$ producing $\Delta^{(2)}(0,1)$ and $\Delta^{(2)}(1,1)$ as
in figure \ref{fig1}. The next level of segments is again obtained by
contracting the existing segments by one of the four factors. At each
level the parent segment
$\Delta^{(n)}(\varepsilon_{n},\ldots,\varepsilon_{1})$ produces two new
segments at the ratio
\[
	\frac{%
	| \Delta^{(n+1)}(\varepsilon_{n+1},\ldots,\varepsilon_{1}) |
	}{%
	| \Delta^{(n)}(\varepsilon_{n},\ldots,\varepsilon_{1}) |
	} = \sigma_{\varepsilon_{n+1},\varepsilon_{n}}
\]
with the convention that $\varepsilon = 0$ is the left
contraction
and that $\varepsilon = 1$ is the right contraction. This construction can
be generalized by letting the child segments depend on more than
the last two bits, as is the case in the period doubling
scenario. We can also allow certain values of the scaling
function to be one, as long as every segment goes to zero length
as the number of levels goes to infinity.  The segments produced
by this method are naturally organized in a binary tree, as each
parent segment gives rise to two child segments.

Taking the limit of this refinement process produces the fractal.
We are not restricted in this description to one (topological)
dimension: we could have used squares that get divided into
smaller squares instead. In general the scaling function that
describes a fractal is not unique, depending on how the segments
are labeled.

The distinction between fractals and multifractals comes when we
describe them in the way of Hutchinson \cite{Hutchinson}. This
is a characterization of fractals through a dynamics that
generates them.  A fractal $\cal F$ is a compact set of points of
the Euclidean space that is invariant under a finite set of $Q$
contractions $\{ \omega_{i} \}$.  The set $\cal F$ satisfies 
\[
	{\cal F} = \bigcup_{1 \leq i \leq Q} \omega_{i}({\cal F})
\]
with the condition that the images of different $\omega_{i}$ are
contained in opens that do not overlap.  The last condition may
be relaxed to the intersection of images being a countable
set or set of zero Hausdorff measure. The slowest contraction rate
of
a transformation $\omega_{k}$ is its contraction rate
$s_{k}$. If we measure distances in the Euclidean space with 
$d(\cdot \: , \cdot)$, then
for every pair of points $x$ and $y$ of the fractal $\cal F$, the
number $s_{k}$ satisfies 
\begin{equation}
 \begin{array}{c}
	d( \omega_{k}(x) , \omega_{k}(y) ) \leq
	  s_{k} d(x,y) \\ 0 < s_{k} < 1 
 \end{array}
 \label{contraction}
 \mbox{ .} 
\end{equation} 
To fix a particular $s_{k}$ we choose the minimum of all numbers
that satisfy the inequalities.
If the $\omega_{i}$ are uniform contractions, {\em i.e.\/}, in
(\ref{contraction}) we have an equality instead of an
inequality, the fractal $\cal F$ is uniformly
self-similar. In this case it can be shown that the contractions
are the composition of a rotation, a translation and a uniform
contraction by the rate $s_{k}$ (see Hutchinson
\cite{Hutchinson}). For uniformly self-similar fractals one has the result:

{\em Theorem} (Hutchinson) If $\cal F$ is uniformly self-similar
then the capacity and the Hausdorff dimension are numerically
equal. In this case the dimension $D$ is the unique root of the
equation
\[
	s_{1}^{D} + s_{2}^{D} + \cdots + s_{Q}^{D} = 1
	\mper
\]
See \cite{Hutchinson} for the proof and a partial converse. 
Many more details on uniformly
self-similar fractals can be found in the book by Barnsley
\cite{BarnsleyFractalsEverywhere}.

Most fractals that arise from dynamical systems are
not uniformly self similar and therefore multifractal, but still it is
conjectured \cite{FarmerDimension,TakensReconstruction,ChicagoFofa}
that the capacity and the Hausdorff
dimension are numerically equal. The Feigenbaum set, the Arnold
tongue structure of the sine circle map, and the orbit of the
irrational winding number of a critical circle map are all
examples of multifractals.

To show that a four scale Cantor set is a multifractal, and 
therefore Hutchinson's theorem does not apply, we consider 
the segments that cover the set,
\(  \Delta^{(n)}(\varepsilon_{n}, \ldots , \varepsilon_{1})  \),
and how they can be transformed into each other. 
The segments are organized by levels, there
being $2^{n}$ segments at level $n$. All the segments at level
$n$ in one region of the set can be expressed in terms of the
segments of another region one level above, {\em i.e.\/}, 
\[
	| \Delta^{(n)}(\varepsilon_{n}, \ldots , \varepsilon_{1}) | =
	| \Delta^{(n-1)}(\varepsilon_{n}, \ldots , \varepsilon_{2}) |
	\sigma_{ \varepsilon_{2} , \varepsilon_{1} }
	\frac
	   { | \Delta^{(1)} (\varepsilon_{1}) | }
	   { | \Delta^{(1)} (\varepsilon_{2}) | }
	\mper
\]
\begin{figure}
  \centerline{\psfig{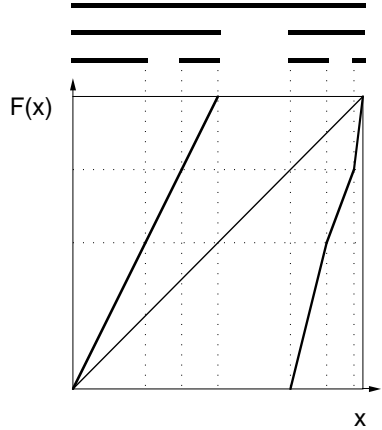}}
  \caption{\protect\footnotesize Presentation function for the four
scale Cantor set.  The first few stages of the construction and
how it relates to the slopes is indicated over the plot.}
  \label{fig2}
\end{figure}
As this relation holds for any $n$, it will hold for
a region as small as needed and therefore for all points of the fractal. As $\varepsilon_{1}$ and $\varepsilon_{2}$ can assume two
values each, there will be four such relations and we
conveniently arrange them into a function $F(\cdot)$ as
indicated in figure \ref{fig2}. We have chosen the scalings to be 1/2,
1/4, 1/2 and 1/8 and the two initial segments to be 1/2 and 1/4
of the unit interval. It is actually the inverse of the slopes
that are plotted, and the function has been arbitrarily defined to
be linear and continuous in the intervals $[1/4,3/8]$
and $[7/8,31/32 ]$. The four scale Cantor set will then be
the set of points that remain in the interval under forward iterations of
$F$. The function $F$ is a presentation function for the four
scale Cantor set, and its inverse has two branches:
\[
     \begin{array}{lr}
	\omega_{1}(x) = F^{-1}(x) &  x \in \Delta^{1}(0) \\
	\omega_{2}(x) = F^{-1}(x) &  x \in \Delta^{1}(1)
     \end{array}
     \mcomma
\]
the contractions to be used in Hutchinson's
characterization of fractals. As $\omega_{1}$ and $\omega_{2}$
are non-uniform contractions, the fractal is not uniformly
self-similar.  The presentation function is not unique, but one
can show, by considering all other ways of organizing the slopes,
that any other presentation function yields a set of non-uniform
contractions. From figure \ref{fig2} one can see how to construct a
presentation function given the first steps generated by the
scaling function. This procedure allows us to go from the scaling function description to the presentation function description as required. The alternative description of multifractals by scaling functions was introduced by Feigenbaum \cite{FeigenbaumScaling}. He later on showed that fractals could also be described with presentation functions \cite{FeigenbaumPresentation,FeigenbaumCirclePresentation}.

Given the scaling function, or the presentation function, or the
the set of contractions $\omega_{i}$, one can compute the
Hausdorff and capacity dimensions of the generated fractal.
To determine the Hausdorff dimension we first consider the 
$\beta$-Hausdorff measure $Z(\beta)$ of the set, computed
from its coverings $\Delta$ and defined as
\[
    Z(\beta) = \lim_{ l \rightarrow 0}
		\inf_{ |\Delta| <  l }
		\sum_{\varepsilon} | \Delta(\varepsilon) |^{\beta}
\]
The infimum is over all countable covers with segments smaller
than $ l$ and the sum is over the segments of the cover. The
measure of the set is zero or infinity for almost all values of
$\beta$, and there is a
unique $\beta$, the Hausdorff dimension $\beta_{H}$, that
separates these two domains of $Z$. Numerically the infimun is
difficult to determine, and it is convenient to consider the
covers from a grid of boxes of side $ l$, so there is no need to optimize the covers by taking the infimum.  The capacity dimension
will be the number that separates the zero and the infinity
domain of $Z$, just like in the Hausdorff case. In the capacity
case all terms of the sum are the same and if $N( l)$
boxes are needed to cover the set, the capacity dimension $\beta_C$
is written as
\begin{equation}
	\beta_{C} = - \lim_{ l \rightarrow 0}
		\frac 
		 {\ln N( l) }
		 {\ln  l }
	\mcomma
	\label{1B}
\end{equation}
when the limit exists.

If the segments of the covers were all
generated from a (translational invariant) scaling function we
can identify $Z$ with the partition function of a thermodynamical
system with energy $- \ln | \Delta(\varepsilon) | $. For the four scale
Cantor set we can write 
\[
	| \Delta^{(n)} ( \varepsilon_{n}, \ldots ,\varepsilon_{1}) | =
	\sigma_{ \varepsilon_{n},\varepsilon_{n-1} }
	\ldots
	\sigma_{ \varepsilon_{3},\varepsilon_{2} }
	\sigma_{ \varepsilon_{2},\varepsilon_{1} }
	| \Delta^{(1)} (\varepsilon_{1}) |
	\stackrel{\rm def}{=}
	\exp ( - u(\varepsilon_{n}, \ldots , \varepsilon_{1} ) )
\]
and 
\[
   Z^{(n)}(\beta) = \sum_{\varepsilon} \exp ( - \beta
	u^{(n)}(\varepsilon) )
\]

To complete the identification we must take the limit 
$n \rightarrow \infty$.
From the partition function we can compute the pressure at
inverse temperature $\beta$ in the thermodynamic limit from
\begin{equation}
    - p(\beta) = \lim_{n \rightarrow \infty} 
	 	\frac{1}{n} 
		\log ( \sum |\Delta^{(n)}(\varepsilon)|^{\beta} )
   \mper
   \label{1C}
\end{equation}
The name pressure is used rather than free energy per particle
per temperature degree because the limit of the logarithm of the partition 
function can be identified with the pressure
in the case of a lattice gas in the grand-canonical ensemble.
Vul, Sinai and Khanin \cite{SinaiReview} and Ruelle
\cite{RuelleBRFormula} have shown that the Hausdorff dimension
$\beta_{H}$ is the unique number for which the pressure satisfies
the Bowen-Ruelle formula
\[
	p(\beta_{H}) = 0
	\mper
\]
Using transfer matrix techniques to evaluate the partition
function, Feigenbaum \cite{FeigenbaumStrangeSets} has shown
that the Hausdorff dimension of
the four scale Cantor set is the root of
\[
	\frac{
	  \sigma^{\beta}_{00} + \sigma^{\beta}_{11} +
	  \sqrt{ ( \sigma^{\beta}_{00} + \sigma^{\beta}_{11})^{2}
	         - 4 \sigma^{\beta}_{01} \sigma^{\beta}_{10} }
	}
	{2} = 1
	\mper
\]
For the scaling coefficients we are considering the root is
$ \beta_{H} = 0.640602307$.

We can also determine the capacity of the fractal from the
partial coverings given by the segments.  To determine the
capacity of the set we must count how many boxes
(segments) of a given size are needed to cover the set.  At level
one we would need two boxes of size 1/2 to cover the set, while if
we had used a box size of 1/8, apparently six boxes would be
necessary at level one. This is an overestimate, as further refinement
of the cover shows that only five boxes are required.  One should
be careful not to confuse the covers of the Cantor set with the
set itself.  The boxes of the grid must cover the set and not one
of its covers.  To count how many boxes of a given size cover the
set, each sub-cover should be refined until it is smaller than the
current box size, and the final number of boxes counted.
\begin{figure}
   \centerline{\psfig{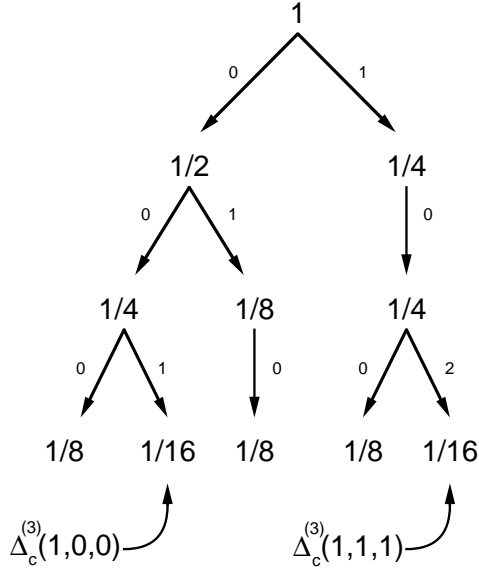}}
   \caption{\protect\footnotesize Labeling scheme for the capacity
   tree.  The branch labels are explained in section 2 and the
   segment labels in section 3.}
   \label{fig4}
\end{figure}

This suggest another organization of the fractal. Instead
of contracting every segment by one of the values of the scaling
function, we fix a box contraction rate per level and at each
level contract the covers that are larger than the
current box size. As the box size goes to zero with increasing
levels of the construction, the final set will the same. As the
slowest contraction rate is 1/2 we will reduce the box size by a
factor of two at each level. This means that the segment that
gets scaled by the factor of 1/8 only needs to be contracted at
every three levels, while the segment that gets scaled by 1/4,
at every two levels, and the segment that gets scaled by 1/2, at
every level.  We now introduce another symbolic labeling for the
segments: each segment gets a 0, 1, or 2 according to how many
levels it can go without contracting. As we started with a box
size of one and took a 1/2, 1/4 contraction to generate the two
initial segments, they can also be labeled according to the new
scheme. In figure \ref{fig4} we have the the first few levels of
coverings, with the labels for counting indicated on the arrows
and two segments labeled as described in section 3.

At each level, if a segment has a label 0 it produces two
daughters, and if it has a label bigger than zero it copies
itself at the next level, but with a label one unit smaller.
If we read the branches of the tree as words built from the 
alphabet 0, 1, 2 the rules for the construction of all the branches
at one level are:
\[
 \begin{array}{rcl}
	x2 & \mapsto & x21 \\
	x1 & \mapsto & x10 \\
	x10 & \mapsto & x100,x102 \\
	x00 & \mapsto & x000,x001
 \end{array}
\]
where $x$ stands for any partial word. One can see by inspection
that these are the rules used for constructing the tree in figure
\ref{fig4}. With the use of these rules we can write relations between the
number of leaves (or words) of one kind with the number
of leaves at previous levels. If $N_{0}(n)$ denotes the number of
segments that have the label 0 at level $n$, and $N_{1}(n)$ and
$N_{2}(n)$ the number of segments with the other labels, we have
that
\[
 \begin{array}{rcl}
	N_{0}(n) & = & N_{0}(n-1) + N_{1}(n-1)  \\
	N_{1}(n) & = & N_{0}(n-1) - N_{1}(n-2) + N_{2}(n-1) \\
	N_{2}(n) & = & N_{1}(n-2)
 \end{array}
\]
with the initial conditions
\[
 \begin{array}{c}
   N_{1}(0) = N_{2}(0) = N_{2}(1) = 0 \\
   N_{0}(0) = N_{0}(1) = N_{1}(1) = 1 
 \end{array}
 \mper
\]
To solve the system of recurrences we use the generating function
method. We will solve for $N_{0}$ as zeros dominate the tree.
Let $\Gamma_{0}(x)$ be the generating function for the the
number of segments with label 0 at level $n$,
\[
	\Gamma_{0}(x) = \sum_{n \geq 0} N_{0}(n) x^{n}
	\mper
\]
By solving the system of recurrences for $N_{0}$ we get
\[
	N_{0}(n) - 2 N_{0}(n+1) - N_{0}(n+3) + N_{0}(n+4) = 0
	\mper
\]
In multiplying by $x^{n}$, summing for $n \geq 0$ , and
subtracting a few initial terms, we get that 
\begin{equation}
	\Gamma_{0} (x) = - \frac{x^{3} - x^{2} -1}
			     {x^{4} - 2 x^{3} - x + 1}
	\mper
	\label{EqG0}
\end{equation}
The same procedure can be used for obtaining the generating
functions for the number of segments with label 1, $\Gamma_{1}$,
and with label 2, $\Gamma_{2}$.  The generating function for
the total number of boxes at each level is given by the sum of
all three generating functions:
\[
	\Gamma (x)  =  \Gamma_{0}(x) + \Gamma_{1}(x) +
		\Gamma_{2}(x) =
		\frac{x^{2}+x+1}{x^{4}-2x^{3}-x+1} =
		\sum_{n \geq 0 } N(n) x^{n}
 \mper
\]

The generating function can be expanded in a Taylor series
around zero and the term in $x^{n}$ picked out to give $N(n)$.
This can be done by factoring the generating function in partial
fractions. Each linear term is then expanded according to the
relation
\[
	\frac{1}{1-ax} = \sum_{n \geq 0} (ax)^{n}
\]
and we see that the leading behavior of the term in $x^{n}$ is
of the form
\[
	N(n) \rightarrow (\mbox{const}) (1/r_{1})^{n} 
\]
as $ n \rightarrow \infty$ and 
where $r_{1}$ is the smallest root of the denominator in absolute
(\ref{EqG0}) value. The roots of the polynomial are given in
table \ref{Roots}.

\begin{table}[bt]
  \begin{center}
  \def\baselinestrech{2}
	\begin{tabular}{|cc|} 
		\hline \hline
		root  & value \\ \hline
		$r_{1}$ & $0.641\, 445$ \\
		$r_{2}$,$r_{3}$  &  $0.858 \, 004 e^{\pm i 0.458 \, 255}$ \\
		$r_{4}$ & $2.177 \, 689$ \\ \hline
	\end{tabular}
  \def\baselinestrech{1}
  \end{center}
  \caption{\protect\footnotesize Poles of the generating function}
  \label{Roots}
\end{table}   
	
The exact capacity of the four-scale Cantor set can be determined
from the asymptotic behavior. We get that
\[
  \beta_{C} = - \frac{\ln r_{1}}{\ln 2} = 0.640602307
  \mper
\]
in complete agreement with the Hausdorff dimension determined by
the thermodynamical formalism. We can verify that $\beta_{C}$ and
$\beta_{H}$ are the same number by verifying that they satisfy
the same algebraic equation.

The above expression for the number of boxes, $N(n)$, needed to cover the Cantor set is an exact expression for a multiscale Cantor set and it is qualitatively different from the expression for the
number of boxes for a uniform scale Cantor set.

\section{The equality of Hausdorff and capacity dimensions}

We now want to show that the capacity and the Hausdorff
dimensions are the same for a multiscale Cantor set. We will
only consider multiscale Cantor sets that are described by a
Feigenbaum scaling function on a binary tree (like the period doubling
tree). These Cantor sets not only have the compact description
offered by scaling function, but they also have a finite
Hausdorff measure (see Hutchinson \cite{Hutchinson}). We can
easily generalize to trees that have a bounded number of branches
per node.

When the scaling function is of infinite range an infinite amount
of initial segments are needed to produce the first level of
refinement. To avoid having to specify an infinite amount of
segments we choose an approximation scheme to generate initial
segments. It will turn out that the thermodynamic quantities of the
set are independent of this scheme. At level $n$ we want to compute
\[
	Z^{(n)}(\beta) = \sum_{\varepsilon}
                  |\Delta^{(n)}(\varepsilon)|^{\beta}
\]
with $\varepsilon$ running over all possible configurations. We 
approximate 
\begin{eqnarray*}
  \lefteqn{|\Delta^{(n)}(\varepsilon_{n}, \ldots ,\varepsilon_{1})| = } \\
  & &
  \sigma(\varepsilon_{n},\ldots,\varepsilon_{1},0,0,\ldots)
  \sigma(\varepsilon_{n-1},\ldots,\varepsilon_{1},0,\ldots)
  \ldots
  \sigma(\varepsilon_{1},0,\ldots)
  |\Delta_{0}|
\end{eqnarray*}
by padding the scaling function with zeros. For the uniqueness of
the thermodynamic quantities 
the variation of the scaling function decreases for bits to the 
right,
\begin{equation}
	|\sigma(\ldots,\varepsilon_n=1,\ldots) -
	  \sigma(\ldots,\varepsilon_n=0, \ldots)| < f(n)
	\label{3f}
\end{equation}
where $f(n)$ goes to zero faster than $1/n^2$ as $n$ goes to
infinity. This is equivalent to 
the uniqueness of the Gibbs state (see Simon's lemma \cite{SimonLemma}).
A consequence of the
uniqueness is that the thermodynamic properties are independent
of the boundary condition, and for the purpose of computing
these properties almost any choice of boundary will do, and that
is why we need not specify an infinite number of initial
conditions. 

To compute the capacity we must evaluate its defining limit (\ref{1B}).
The function $N(l)$ is piecewise constant, as it takes values over
the integers, and monotonically increasing, as a refinement
cannot decrease the number of boxes required to cover the set.
If the capacity of the set is well
defined, then we can choose any sequence $l_{n}$ of box sizes as
long as $l_{n} \rightarrow 0$ as $ n \rightarrow \infty$, and
evaluate the limit of the sequence of ratios
\[
	\beta_{C} =
	- \lim_{l \rightarrow 0} \frac{ \ln N(l) }{\ln l}
	= 
	- \lim_{ n \rightarrow \infty} \frac{ \ln N(l_{n})}{ \ln l_{n}}
	\mper
\]

The covers $\Delta^{(n)}(\varepsilon)$ can be converted into a 
new sequence of covers as long as we still have the convergence
to zero box size.
We fix the contraction rate $b$ of the grid of boxes to be the slowest
contraction rate of any child segment, that is,
\[
        b= \sup_{\varepsilon} \; \sigma(\varepsilon) < 1
        \mper
\] 
If we denote the new covers by $\Delta^{(n)}_{C}(\varepsilon)$ then they
will satisfy
\begin{equation}
        c b^{n} | \Delta_{0} | \leq
        | \Delta^{n}_{C}(\varepsilon) |  \leq
        b^{n} | \Delta_{0}|
        \label{eq5}
\end{equation}
where $c$ is the non-zero infimum of the scaling function over
all configurations $\varepsilon$. This can be seen as tree where the
segments contract only if at the next level they were to be
larger than the box size at that level. This tree --- the
capacity tree --- allows us to define a new scaling function.

The capacity tree is a different set of covers that converge to the
same set. Each point $x$ of the fractal can be uniquely defined
by the set of segments that contain it,
\[
        \Delta_{0} \supset
        \Delta_C^{(1)}(\varepsilon_{1}) \supset
        \Delta_C^{(2)}(\varepsilon_{2},\varepsilon_{1}) \supset
        \ldots
        \ni x
\]
and the sequence of $\varepsilon_{i}$ are the unique configuration
associated with the point $x$ and they are used to label the covers
as done in figure \ref{fig4}. This uniqueness establishes a one-to-one
correspondence between the limit points of the capacity tree and
the Hausdorff tree.   It is important to realize that the points
of the Cantor set do not change when we go from the Hausdorff tree
to the capacity tree, only the sequence of covers that approach
them.   The sequence of covers approaching a point in the Hausdorff
tree and the capacity tree do not differ, only their labels do.

We can now use the new sequence of covers $\Delta^{(n)}_{C}(\varepsilon)$
in formula (\ref{1C}) to compute the Hausdorff dimension.  As the
set is the same, the Hausdorff dimension will not change.  This
will be the case if the original scaling function fell off
exponentially, for the transformation to the capacity tree just
changes the rate of fall off of $f(n)$ in (\ref{3f}).  But if the
original scaling function were bounded by power law fall off, $f(n)=
n^{-p}$, then there will be cases when the re-arrangement cannot
be done even with a unique Gibbs state.

To compute the capacity we rewrite the limit (\ref{1B}) in terms
of thermodynamical quantities.  $N(n)$ is the number of segments
at the level $n$, and by the construction of the capacity tree
\[
        N(n)  =  \mbox{card} \{
          \Delta^{(n)}(\varepsilon) :  c b^{n}|\Delta_{0}| \leq
                |\Delta^{(n)}_{C}(\varepsilon)| \leq
                b^{n} |\Delta_{0} \}
        \mper
\]
The bounds for $|\Delta^{(n)}_{C}(\varepsilon)|$, equation (\ref{eq5}),
tell us that $(\ln |\Delta^{(n)}_{C}| )/n$ is approaching a limit as
$n \rightarrow \infty$. It differs from this limit by at most
$ -( \ln c |\Delta_{0}| )/n $, a small quantity we shall call
$\delta_{n}$. The number of occupied boxes at level $n$ can be
written as
\[
        N(n) =
        \mbox{card}
          \{ \Delta^{(n)}(\varepsilon) :
            | \frac{1}{n} \ln |\Delta^{(n)_{C}}(\varepsilon) | - \ln b |
           \leq \delta_{n} \}
        \mper
\]
By comparing the expression for $N(n)$ with the definition of
entropy in the microcanonical (see the lectures by Lanford
\cite{LanfordLectures}), we see that in the limit $n$ goes to
infinity $(\ln N(n) )/n $ is the entropy of the states with
energy $\ln b^{-1}$. Also $l_{n}$ can be expressed in terms of
$\ln b$. Combining these results the capacity can be expressed as
the ratio of the entropy to the energy in the microcanonical
ensemble,
\begin{equation}
        \beta_{C} =
        \frac
         {\displaystyle \lim_{n \rightarrow \infty} \frac{1}{n} \ln N(n) }
         {\displaystyle \lim_{n \rightarrow \infty} \frac{1}{n} \ln l_{n} }
        =
        \frac{s(u)}{u}
        \mper
        \label{eq4}
\end{equation}

The capacity tree corresponds to a statistical mechanical system
where all the states have the same energy $u$. If the energy can only
assume one value, then the entropy is also only defined at that single
point. To relate this to the canonical description
we have to Legendre transform the energy to get the free energy,
or alternatively, transform the entropy to get the pressure. In
this case we cannot compute the derivative of the entropy to get
the inverse temperature $\beta$ of the system because the entropy
is only defined at a point. But the thermodynamical formalism is
more general, and the Legendre transform can be performed just
with continuity.

If $f(x)$ is a convex function its Legendre transform is given by
the following procedure: pick any real number $p$ and consider
the real line $y=xp$; the maximum signed distance (not including
infinity) between the curve $f(x)$ and the line $y=xp$ along the
vertical is the Legendre transform of $f$ for the value $p$. This
reduces to the usual definition when $f$ is differentiable.

The Legendre transform of the entropy is then a line. If $s_{0}$
is the unique value of the entropy and $u_{0}$ the unique value
of the energy, the pressure as a function of Legendre variable
$\beta$ is 
\[
	p(\beta) = \beta u_{0} - s_{0}
	\mper
\]
For the Hausdorff dimension, $\beta_{H}$, we have that
$p(\beta_{H})=0$ and, as both computations are for the same set
of states,
\[
	\beta_{H} = \frac{s_{0}}{u_{0}} = \beta_{C}
\]
according to equation (\ref{eq4}), establishing the equality of
the two dimensions.

\section{Conclusion}

With the aid of the thermodynamic formalism we have shown (within
the precision of a physicist) that for a fractal described by a
scaling function that falls off exponentially, the capacity and
the Hausdorff dimension will be numerically equal.  If the fractal
is generated from the iterations of a differentiable presentation
function, then the scaling function will fall off exponentially
fast, the rate related to its Holder exponent.  The equality may
hold in less restrictive conditions, as we only need the existence
and uniqueness of the ratio (\ref{eq4}) between the entropy and
the energy. Even when there is more than one Gibbs state the ratio
may be the same in each state due to a symmetry of the scaling
function. The ease in which the result was established is due to
the use of the scaling functions. They play the role of the
interaction in the thermodynamics of iterated functions.  If instead
we had started with the derivative of the iterating map, the analysis
would become more difficult.

Not every map leads to an exponentially decaying scaling function.  The most common exception is for maps that display intermittency.   In this case the scaling function will have a power law decay --- in equation (\ref{3f}),  $f(n) = n^{-s}$.   In most cases $s<2$ indicating that the generalized dimensions of the set have a phase transition.   For this case the Hausdorff tree and the capacity tree are not equivalent, as they cannot be re-arranged into each other. 

Obtaining the scaling function from the map, and proving it
unique, has only been done in a few cases.  One can intuitively
see the difficulties by noticing that when we constructed the
presentation function for the four scale Cantor set, we had the
freedom to define the map arbitrarily in the pre-image of the
gap, {\em i.e.\/}, for $F^{-1}( [1/2,3/4] )$.
For any choice in the gap pre-image we would had obtained
the same scaling function and the same limit set. This
arbitrariness extends to all pre-images of the gap, that is, the
presentation function is arbitrary for most points.  It is the
product of the slopes of the presentation function at the points
of the limit set that determine the scaling function, a property
that seems difficult to control with a priori bounds. If however
we start our analysis from the scaling function or the
presentation function we can easily determine the thermodynamic
quantities from them.

\section*{Acknowledgments}

I would like to thank John Lowenstein and Alan Sokal for critical
discussions.

\bibliographystyle{unsrt}

\end{document}